\documentclass[a4paper,11pt]{article}
\usepackage{jheppub} 
\usepackage{lineno}
\usepackage{graphicx} 
\usepackage{amsfonts}
\usepackage{dsfont}
\usepackage{amsmath}
\usepackage{physics}
\usepackage{hyperref}
\usepackage{xcolor}
\usepackage{mdframed}
\usepackage{soul}

\newcommand{\be}{\begin{equation}}
\newcommand{\ee}{\end{equation}}

\newcommand{\ba}{\begin{aligned}}
\newcommand{\ea}{\end{aligned}}

\definecolor{petrolblue}{HTML}{00728D}
\definecolor{darkgray}{HTML}{4D4E4F}
\definecolor{white}{HTML}{FFFFFF}
\definecolor{darkblue}{HTML}{183F56}
\definecolor{lightgray}{HTML}{C8CAD4}
\definecolor{lightblue}{HTML}{9CD7F3}
\definecolor{red}{HTML}{AB3502}
\definecolor{purple}{HTML}{4B70FF}
\definecolor{orange}{HTML}{E69426}

\newcommand{\dint}{\text{d}}

\def \manspacing {.157cm}

\title{Bananas are Unparticles: \\ Differential Equations and Cosmological Bootstrap}

\author[a]{Tom Westerdijk}
\emailAdd{tom.westerdijk@sns.it}

\author[a]{, Chen Yang}
\affiliation[a]{Scuola Normale Superiore and INFN, \\ Piazza dei Cavalieri 7, Pisa, 56126, Italy}
\emailAdd{chen.yang@sns.it}

\abstract{
We establish an exact correspondence between tree-level cosmological correlators with unparticle exchange (at integer scaling dimensions) and banana diagrams of conformally coupled scalars. This duality enables us to systematically solve the governing differential equations through the application of shift relations and cosmological bootstrap techniques. Furthermore, we adapt a dimensional regularization scheme to cosmological correlators, demonstrating how renormalization conditions uniquely fix the regularization prescription. Our results provide new insights into the analytic structure of higher-order loop corrections to inflationary correlation functions. 
}

\begin{document}

\maketitle

\flushbottom

\section{Introduction}
\label{sec:intro}
A striking disparity persists between our understanding of perturbative quantum field theory (QFT) in cosmological spacetime and its well-established Minkowski counterpart. 
One interesting direction of investigation is to push beyond tree-level computations. Computing loop diagrams is a notoriously thorny subject in theoretical cosmology, for which a systematic theory (even at one loop) remains elusive. 
Another example of unexplored territory is to consider strongly coupled sectors that are coupled to the inflaton \cite{Pimentel:2025rds}. 
It is possible that in the primordial universe, in addition to the Standard Model, the ultraviolet (UV) theory contains sectors that become non-trivially scale-invariant at energies comparable to the Hubble scale during inflation. 
This has potentially interesting phenomenological consequences, discussed in \cite{Green:2013rd,Kikuchi:2007az,Pimentel:2025rds}, and has not been explored with the same intensity as the case of weakly coupled models. 
In particle physics, this non-trivial scale-invariant sector can be observed experimentally by measuring missing energy distributions, for example in \cite{Fox:2011pm,CMS:2014jvv,Sannino:2009za,Cheung:2007zza,Cheung:2007ap}. 
There is a corner where these two subjects touch each other: the correlators due to the exchange of ``unparticles" with integer scaling dimensions are equivalent to a certain class of ``banana" loop integrals. 

\begin{figure}[h]
    \centering
    \includegraphics[width=0.595\linewidth]{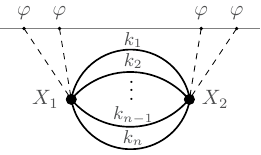}
    \caption{A diagram representing the $(n-1)$-loop banana correlator}
    \label{fig:banana}
\end{figure}

The weakly coupled sector in cosmology can be characterized by its correlators, computed in perturbation theory. 
Although a complete understanding of tree-level is well underway, almost nothing is known at loop-level. 
As usual, the computation can be organized in terms of Feynman diagrams. 
Focusing on diagrams with two interaction vertices, a general class of loop-corrections have the topologies illustrated in Figure \ref{fig:banana}. 
These are called \textit{banana} loops and have been extensively studied in the amplitudes literature \cite{Groote:2005ay,Bzowski:2013sza,Mishnyakov:2023wpd,Cacciatori:2023tzp,Duhr:2025ppd,Duhr:2025tdf}; in the cosmological context, only a handful of \textit{analytic} expressions exist for the special case of de Sitter space \cite{Chowdhury:2023arc,Cacciatori:2024zrv, Cacciatori:2024zbe,Qin:2023nhv,Qin:2024gtr,Benincasa:2024ptf}.
Although there are detailed analytic results for the 1-loop banana, i.e. the bubble, our knowledge about multi-loop correlators is rather limited.
\vspace{-.1em}

One of the confusing aspects of loops in cosmology is the interplay between their UV and infrared (IR) divergences \cite{Maldacena:2002vr,Senatore:2009cf,Senatore:2012nq,Pimentel:2012tw,Arkani-Hamed:2015bza,Anninos:2014lwa,Salcedo:2022aal,Cespedes:2023aal,Baumgart:2019clc,Gorbenko:2019rza}, requiring regularization and proper renormalization. 
We will show with an explicit example that consistent renormalization with local counterterms requires a regularization scheme that mixes the UV and IR-effects in a particular way. 
Interestingly, the same divergences appear in the study of unparticles\footnote{Among the possibilities for long-distance QFTs, there are theories with non-trivial anomalous dimensions. 
From a particle physics perspective, the resulting scale-invariant field is called an \textit{unparticle}~\cite{Banks:1981nn,Georgi:2007ek,Georgi:2007si,Georgi:2009xq,Grinstein:2008qk,Stephanov:2007ry}. In order to capture some general features of the scaling invariant sector, it is reasonable to regard the unparticles as conformal fields.} --- where renormalization is needed to define the composite operators with integer scaling dimensions. 

A physically motivated idea to tackle the multi-loop banana integrals would be to bundle all $n$ internal legs together and regard them as a single, composite particle propagating between the two interaction sites. 
If the particles running through the loops are all identical conformally coupled scalar fields, the intermediate single state exactly corresponds to an unparticle with integer (four-dimensional) scaling dimension.\footnote{This is not to be confused with the scaling behavior of quantum fields in their conformal boundary, which is what one usually refers to when discussing holographic descriptions of cosmology.}

This relation is a special case of the spectral decomposition, a powerful relation successfully used to compute loop corrections in (Anti-)de Sitter spacetime (see e.g. \cite{Hogervorst:2021uvp,DiPietro:2021sjt,Loparco:2023rug,Marolf:2010zp,Fitzpatrick:2011hu}). The spectral decomposition 
in (Anti-)de Sitter expresses 
any two point function as a sum or integral of propagators of conformal primaries.
In particular, two-point functions of conformally coupled scalar composites can be rewritten in terms of propagators of unparticles with integer conformal weight. 
We demonstrate that this equivalence  readily generalizes to any power-law cosmology. 
Studying unparticle-exchange correlators in this broad class of cosmological backgrounds, we develop a bootstrap-inspired method for deriving loop corrections to cosmological correlators. Moreover, it allows us to concretely show how de Sitter symmetries prefer a specific regulator for the loop integrals.

\subsection*{Outline}
\vspace{\manspacing}

In Section \ref{sec:bananas_are_unparticles}, we will show that the integral computing a banana loop can be recast into the integral representation for the unparticle exchange correlator. 
We will exploit this equivalence to derive the differential equations for the $(n-1)$-loop banana correlator in any power-law cosmology in Section \ref{sec:DE_eqns}.
Employing a rudimentary ``cosmological bootstrap," we derive an analytic expression for the bubble from a minimal set of differential equations and a handful of physical constraints in Section \ref{sec:seed}. 
This bubble will serve as the seed to build the whole tower of $(n-1)$-loops with shift relations as is done in Section \ref{sec:shift}. 
Finally, in Section \ref{sec:renorm}, we will discuss the subtleties of renormalization using local counterterms in an expanding spacetime and see how it is intricately connected to the type of regularization we have to use. 
We conclude and mention some directions of future work in Section \ref{sec:discussion}. 
The Appendix \ref{app:shift_operator} shows an explicit derivation for the shift operator used in the main text. 

\subsection*{Setup}
\vspace{\manspacing}

The general action of a single scalar field $\phi$ in a fixed $D$-dimensional curved spacetime with polynomial interactions can be written as 
\begin{align}
    S = -\int \dint^Dx \sqrt{-g} \left(\frac{1}{2}\partial_\mu\phi\partial^\mu\phi + \frac{1}{2}m^2\phi^2 + \sum_n \lambda_n\frac{\phi^n}{n!}\right), 
    \label{def:EFT_FRW}
\end{align}
where $\lambda_n$'s are arbitrary small coupling constants. 
The Friedmann-Robertson-Walker (FRW) metric for a power-law cosmology in flat-slicing is given by
\begin{align}
    \dint s^2 = \frac{-\dint \eta^2 + \dint \vec{x}^2}{(\eta/\eta_0)^{2(1+\delta)}} \equiv a(\eta)^2 (-\dint \eta^2 + \dint \vec{x}^2). 
    \label{def:FRW_metric}
\end{align}
Throughout the paper, we work with $\eta_0\equiv-1$. 
We regularize the UV-divergences of the loops with dimensional regularization (dim-reg): $D\equiv d+1=4+2\epsilon$. 
Notice that $\epsilon$ is a complex parameter. 

To be more precise, when we refer to the $n$-th banana loop in this paper, we mean a correlator with two interaction vertices and $n$ internal lines thus forming an $(n-1)$-loop as in Figure \ref{fig:banana}. 
The loops are composed of conformally coupled scalars, $\varphi$, whose mass is generated by a coupling to the Ricci scalar,
\begin{align}
    S_{\varphi} =-\frac{1}{2} \int \dint^D x\ \sqrt{-g} \left(\partial_\mu\varphi\partial^\mu\varphi + \xi R \varphi^2\right), 
\end{align}
with the coupling constant $\xi\equiv \frac{D-2}{4(D-1)}$. 
In the Bunch-Davies vacuum, the mode function of conformally coupled scalars is
\begin{align}
    f_k(\eta) = (-\eta)^{1+\delta}\frac{e^{-ik\eta}}{\sqrt{2k}}.
\end{align}
Here $k\equiv|\vec{k}|$ is the magnitude of the three-dimensional momentum. 

In the ``In-In formalism", the equal-time four-point, two-site correlator can be divided into four ``sectors'' based on time-ordering: 
\begin{align}
    &\langle\varphi_1\varphi_2\varphi_3\varphi_4\rangle \equiv I_{++} + I_{+-} + I_{-+} + I_{--}, \quad \text{where} \quad I_{--} = I_{++}^*, \quad I_{-+} = I_{+-}^*. 
\end{align}
Diagrammatically, the $I_{++}$ sector is represented by Figure \ref{fig:banana}. 
For a banana loop exchanging $n$-conformally coupled scalars, the $I_{\pm\pm}$ sectors are defined as 
\begin{align}
    \begin{aligned}
        I_{\pm\pm} = \frac{(-\lambda_n^2)\eta_*^{4(1+\delta)}}{16p_1p_2p_3p_4} &\int_{-\infty}^{0} \frac{\dint \eta_1}{(-\eta_1)^{\alpha+1}}\frac{\dint \eta_2}{(-\eta_2)^{\alpha+1}}  e^{\pm i\eta_1 X_1}  e^{\pm i\eta_2 X_2} \\
        \times&\prod_{i=1}^{n} \int \frac{\dint^{d} k_i}{(2\pi)^d} 
        G_{\pm\pm}(k_i;\eta_1,\eta_2)
        \delta^{(d)}(\vec{Y}+\vec{k}_1+\dots+\vec{k}_n), 
        \label{def:banana_int}
    \end{aligned}
\end{align}
where $\vec{Y}$ is the total external three-momentum, $\alpha \equiv (1+\delta) \big(2-
n(1+\epsilon)\big)-1$ and $\eta_*$ denotes the late-time surface where the correlation functions live. 
The external kinematic variables can be compactly written as 
$X_1 \equiv p_1+p_2$, $X_2 \equiv p_3+p_4$ and $\vec{Y}\equiv\vec{p}_1+\vec{p}_2=-\vec{p}_3-\vec{p}_4$. 
Having absorbed the $\eta_1$ and $\eta_2$ power-laws into the measure, the bulk-to-bulk propagators of $\varphi$ take the same form as in flat space,
\begin{align}
    G_{++}(k_i;\eta_1,\eta_2) &= \frac{1}{2 k_i} \left( e^{-i k_i (\eta_1 -\eta_2)} \theta(\eta_1-\eta_2) + e^{-i k_i (\eta_2 - \eta_1)} \theta(\eta_2 - \eta_1) \right), \\
    G_{+-}(k_i;\eta_1,\eta_2) &= \frac{1}{2 k_i} e^{-i k_i (\eta_2 - \eta_1)}. 
\end{align}
We will highlight the key points for the $I_{++}$ sector as defined in \eqref{def:banana_int} and provide the necessary details to generalize our findings to the other sectors. 

\section{Bananas are Unparticles}
\label{sec:bananas_are_unparticles}

In this section, we will start by introducing an integral representation for cosmological correlators with an exchange described by a two-point function of unparticles.
Subsequently, directing our focus on the bananas, we will find a conformal field theory (CFT) two-point function hiding in the loop integrals. Using this fact, we will show the exact equivalence of certain banana loops and unparticle correlators.

\subsection{Unparticles}
\label{sec:unpar}
\vspace{\manspacing}

The Euclidean flat space two-point function for a conformal scalar field with scaling dimension $\Delta$ is completely fixed by symmetry,  
\begin{align}
    \langle \mathcal{O}_\Delta (\tau_1,\vec{x}_1)\mathcal{O}_\Delta (\tau_2,\vec{x}_2) \rangle_{\text{flat}} = \frac{1}{\Big((\tau_1 - \tau_2)^2 + (\vec{x}_1-\vec{x}_2)^2\Big)^\Delta}. 
    \label{eq:CFTflat}
\end{align}
By scaling the flat space metric with a factor $a(\eta)\equiv 1/(-\eta)^{1+\delta}$ we arrive at \eqref{def:FRW_metric}. 
Thus the two-point function of conformal fields in FRW spacetime is obtained by a Wick rotation $\tau \rightarrow i\eta$ and rescaling the operators, $\mathcal{O}_\Delta(x) \rightarrow \tilde{\mathcal{O}}_\Delta(a(\eta) x)=a(\eta)^{-\Delta} \mathcal{O}_\Delta$: 
\begin{align}
    \langle \tilde{\mathcal{O}}_\Delta (\eta_1,\vec{x}) \tilde{\mathcal{O}}_\Delta (\eta_2,\vec{y}) \rangle_{\text{FRW}} &= \frac{(\eta_1\eta_2)^{\Delta(1+\delta)}}{\Big(-(\eta_1 - \eta_2)^2 + (\vec{x}-\vec{y})^2\Big)^\Delta}. 
    \label{eqn:2pt_dS4}
\end{align}
Assuming the unparticle here is a scalar CFT, we need to respect the unitarity bound for scalar conformal fields in $D$ dimensions: $\Delta\geq(D-2)/2$. 
We refer the readers to \cite{Rychkov:2016iqz} for a pedagogical introduction to conformal field theory. 

It is shown in \cite{Pimentel:2025rds} that the integral $I_{++}$, for the $(++)$-branch with unparticle exchange, can be regarded as an integral transform of its flat space conformally coupled scalar counterpart. The procedure starts by deforming the kinematic variables, $X_1\rightarrow X_1+x_1$, $X_2\rightarrow X_2+x_2$, and $Y\rightarrow Yt$, and then integrating over the parameters $x_1$, $x_2$ and $t$: 
\begin{align}
    I_{++} &\propto \int_0^\infty\dint x_1 \dint x_2\ (x_1 x_2)^{\epsilon_1} \int^\infty_1 \dint t \ (t^2-1)^{\epsilon_2}\ I_{++,\Delta=1}^{(\text{flat})}(X_1+x_1,X_2+x_2,Yt) \\
    &\begin{aligned}
        \propto& \int_0^\infty \dint x_1 \dint x_2 \ (x_1 x_2)^{\epsilon_1}  \int^\infty_1 \dint t \ (t^2-1)^{\epsilon_2}\ \frac{1}{x_1+x_2+X_1+X_2} \\
        &\times \left(\frac{1}{x_1+t+X_1}+\frac{1}{x_2+t+X_2}\right), 
    \end{aligned}
    \label{def:twi_int}
\end{align}
where the twisted parameters $\epsilon_1$ and $\epsilon_2$ are defined as 
\begin{align}
    \epsilon_1 = (1+\delta)(2-\Delta)-1, \quad \epsilon_2 = \Delta-2-\epsilon, 
    \label{eq:UnparticleTwists}
\end{align}
and we rescale $x\rightarrow Yx$, $X\rightarrow YX$ in the second step. 

\subsection{The Banana Integral}
\label{sec:banana}
\vspace{\manspacing}
 
Before strapping our boots, we will show the equivalence of banana loops and unparticle exchanges at the level of the integrals for a certain choice of parameters.
Using a Schwinger parametrization for the $\eta$-power laws, 
\begin{align}
    \frac{1}{\eta_1^{\alpha+1}} \propto  \int_0^\infty \dint x_1\ e^{ix_1\eta_1} x_1^{\alpha}, 
\end{align}
the FRW correlator becomes an integral transform of the flat space result with shifted ``vertex energies", $X_1\rightarrow X_1+x_1, X_2\rightarrow X_2+x_2$, in the same manner as in \eqref{def:twi_int} and \cite{Arkani-Hamed:2023kig}: 
\begin{align}
    I_{++} \propto \int_0^\infty\dint x_1 \dint x_2\ (x_1 x_2)^{\alpha} I_{++}^{(\text{flat})}(X_1+x_1,X_2+x_2,Y).
    \label{eq:flatToFRWban}
\end{align}
For $\alpha=-1$, the correlator will reproduce the flat space result. 
By virtue of this simple integral transform, we can focus on computing $I_{++}^{\text{(flat)}}$ first.

The product of bulk-to-bulk propagators contains only two non-zero terms corresponding to the two ways of time-ordering the two vertices. 
Cross-terms vanish as they include $\theta(\eta_1-\eta_2)\theta(\eta_2-\eta_1)$, which is zero in a distributional sense. 
Consequently, the time integrals of \eqref{def:banana_int} give rise to a simple rational function: 
\begin{align}
    \frac{1}{X_1+X_2}\Bigg(\frac{1}{X_1+\sum_{i=1}^n k_i} + \frac{1}{X_2
    +\sum_{i=1}^n k_i}\Bigg). 
\end{align}
The ``total energy" pole, $1/(X_1+X_2)$, has no $\vec k_i$ dependence and will hence appear as only a prefactor for the loop-integral. 
From the expression in brackets, it suffices to consider only a single term since the two terms are related by permuting $X_1 \leftrightarrow X_2$. 
As we will see later on, the calculation is simplified dramatically by a Schwinger-parametrization of this term, 
\begin{align}
    \frac{i}{X_1+\sum_{i=1}^n k_i} = \lim_{\varepsilon\rightarrow0} \int_0^{\infty(1-i \varepsilon)} \dint s\ e^{-i s (X_1+\sum_{i=1}^n k_i)},
\end{align}
which can be thought of as keeping a single time-integral around. The slight rotation of the contour into the lower half complex plane ensures convergence.
Using the integral representation of the delta function, we obtain 
\begin{align}
    \prod_{i=1}^n \int \frac{\dint^{d} k_i}{(2\pi)^d} \delta^{(d)}(\vec{Y}+\vec{k}_1+\dots+\vec{k}_n) = \int \dint^d x\ e^{i\vec{x}\cdot\vec{Y}} \prod_{i=1}^n \int \frac{\dint^{d} k_i}{(2\pi)^d}\ e^{i\vec{x}\cdot\vec{k}_i}.
    \label{def:vol_conv}
\end{align}
To summarize, we find that we can write the $(++)$-sector of the flat space correlator as
\begin{equation}
    I_{++}^{(\text{flat})}(X_1,X_2,Y) = \frac{1}{X_1+X_2} \Big(J(X_1,Y) + J(X_2,Y)\Big)
\label{eq:I++J++}
\end{equation}
which is completely determined by the integral,
\begin{align}
    J(X,Y)\equiv\int_{0}^\infty \dint s\ e^{-i s X} \int \dint^d x\ e^{i\vec{x}\cdot\vec{Y}} \prod_{i=1}^n \int \frac{\dint^{d} k_i}{(2\pi)^d}\ \frac{e^{i\vec{x}\cdot\vec{k}_i-i s k_i}}{2 k_i}. 
    \label{eq:flat1}
\end{align}
Notice that the same integral $J$ also allows us to write down the $(+-)$-sector since the associated rational functions are proportional to
\begin{equation}
    \frac{1}{X_1-X_2} \left( \frac{1}{X_1 + \sum_{i=1}^n k_i} - \frac{1}{X_2 + \sum_{i=1}^n k_i} \right).
\end{equation}
Each individual $k_i$ integral from \eqref{eq:flat1} can be decomposed as
\begin{align}
    \int \hspace{-.3em}\frac{\dint^{d} k_i}{(2\pi)^d}\ \frac{e^{i(\vec{x}\cdot\vec{k}_i-s k_i)}}{2 k_i}= \frac{\mathds{V}_{S^{d-2}}}{2}\hspace{-.3em}\int_0^\infty\hspace{-.5em}\dint k_i \, k_i^{1+2\epsilon}\hspace{-.2em}\int_{-1}^1\hspace{-.3em}\dint u_i \, (1-u_i^2)^\epsilon \, e^{i k_i(-s+x u_i)}, 
\end{align}
where $\mathds{V}_{S^{d-2}}$ is the volume of a $(d-2)$-unit-sphere and $u_i \equiv \cos \theta_i$. 
Thus for each $k_i$, we have 
\begin{align}
    & \int_0^\infty\dint k_i \int_{-1}^1\dint u_i\ k_i^{1+2\epsilon} (1-u_i^2)^{\epsilon} \ e^{i k_i(-s+ u_i x)} \nonumber \\
    =& \ \Gamma(2\epsilon+2) \int_{-1}^1\dint u_i\ (1-u_i^2)^{\epsilon} (i s-iu_ix)^{-2\epsilon-2} \nonumber\\
    =& \ 2^{2 \epsilon +1} \Gamma (\epsilon +1)^2 (-s^2+x^2)^{-\epsilon -1}, 
\end{align}
where the small negative imaginary part of $s$ ensures the validity of the last step.
By means of the factorization in \eqref{eq:flat1}, the integrand of the position space Fourier transform is simply the $n$-th power of the above,
\begin{align}
    J=c_\epsilon\int_{0}^\infty \dint s\ \int \dint^d x\  e^{i(-s X+\vec{x}\cdot\vec{Y})} (-s^2+x^2)^{-n(1+\epsilon)}, 
    \label{eq:BanFlatFinal}
\end{align}
where $c_\epsilon=(2^{2 \epsilon}\mathds{V}_{S^{d-2}}\Gamma(1+\epsilon)^{2})^n$. 
Strikingly, the integrand resembles a CFT 2-point function in Lorentzian signature, an observation which can be made more manifest by denoting $P=(X-i\varepsilon, \vec Y)$, $X=(|s|,\vec x)$ and $\tilde{\Delta}=n(1+\epsilon)$: 
\begin{equation}
    e^{i(-s X+\vec{x}\cdot\vec{Y})} (-s^2+x^2)^{-n(1+\epsilon)} \equiv e^{i P \cdot X} \frac{1}{(X^2-i\varepsilon)^{\tilde{\Delta}}}. 
\end{equation}
Although tempting, we cannot identify $J$ with the four-momentum Fourier transform of the CFT 2-point function due to the lower bound of the $s$-integral being $0$ and not $-\infty$.

Nevertheless, we can proceed in a similar fashion by first Wick rotating $s$, as $s\rightarrow-i\tau$, which gives us 
\begin{align}
    J = -ic_\epsilon \int_{0}^\infty \dint \tau\ e^{-X \tau} \int \dint^d x\  e^{i\vec{x}\cdot\vec{y}} (\tau^2+x^2)^{-\tilde{\Delta}}. 
\end{align}
From \cite{Grinstein:2008qk,Bautista:2019qxj,Pimentel:2025rds}, we know that in $(d+1)$-dimensional Euclidean space, the spatial Fourier transformation can be brought to the form,
\begin{align}
    &\int \dint^{d} x\ e^{i\vec{x}\cdot\vec{Y}}\frac{1}{(\tau^2+x^2)^{\tilde{\Delta}}} \nonumber \\
    =\ &\frac{(2\pi)^{(d+1)/2}}{4^{\tilde{\Delta} - (d+1)/4}\Gamma(\tilde{\Delta})\sqrt{\pi}} \frac{(2Y)^{\tilde{\Delta} - \frac{d}{2}} }{\tau^{\tilde{\Delta} - \frac{d}{2}}} K_{\tilde{\Delta} - \frac{d}{2}}\Big(Y \tau\Big) \nonumber \\
    \propto\ & \int^{\infty}_1\dint t\ (t^2-1)^{\tilde{\Delta}-\frac{d+1}{2}} e^{-Y \tau t}, 
    \label{eqn:crucial_step}
\end{align}
where we have used an integral representation of the Bessel $K$ function in the last step. 
At this point, evaluating the $\tau$-integral yields a simple rational function,
\begin{align}
    \int_{0}^\infty \dint \tau\ e^{-X \tau} e^{-Yt\tau} = \frac{1}{X+Y t}. 
\end{align}
Once we extract $Y$ from the integral and rescale $X\rightarrow Y X$, substituting $J$ into \eqref{eq:I++J++} yields a compact integral representation for the $(n-1)$-banana loop correlator. 

Most strikingly, this integral representation for the $(n-1)$-banana loop is of exactly the same form as the unparticle correlator \eqref{def:twi_int} with a scaling dimension,
\begin{align}
    \Delta = \tilde{\Delta} \equiv n(1+\epsilon).
\end{align}
The $\tilde{\Delta}$ simply counts the number of exchanged conformally coupled scalars, each of which has a conformal weight $(1+\epsilon)$. 
This dictionary can be further generalized to an arbitrary number of external legs on each site. 

An unparticle field $\mathcal{O}_n(x)$ can be regarded as a single \textit{composite operator} $:\varphi(x)^n:$ when its scaling dimension is an integer multiple of the conformally coupled scalars. 
This composite operator is constructed from a product of fields, $\varphi(x_1)\varphi(x_2)\dots\varphi(x_n)$, where we overlap the points in configuration space, $x_1=x_2=\dots=x_n$.  
A detailed introduction to composite operators is given in \cite{Collins:1984xc}. 

The relation to the banana loop is straightforward: the $n$ scalars in the loop all attach to an interaction vertex at the same point in position space, and thus they are nothing more than the composite operator $:\varphi(x)^n:$ for integer $n$.
Using this picture, the banana loop with identical internal lines can be regarded as a tree-level diagram with an exchanged composite operator.
This means that we can interpret the $\Delta$ poles discussed in \cite{Pimentel:2025rds} as loop divergences in general FRW universes.

The perturbative expansion of strongly coupled states is usually complicated, like matching the high-energy QCD states to low-energy hadron states. 
However, if we just consider a perturbative calculation of \eqref{def:EFT_FRW} up to $O(\lambda_n^2)$, the banana diagrams will be the only non-trivial diagram that we can draw based on the simple two-site topology. 
The two-site multi-loop integrals with self-contractions will vanish when we expand near the vacuum because any self-contraction resembles a tadpole. 
For non-vanishing one-point cases, we refer the reader to \cite{Iliesiu:2018fao,Karlsson:2021duj}. 

This lowest order contribution is the limit where the forces binding the constituents of the composite particles become vanishingly weak. 
Therefore, the physics can be described solely by a free theory of the constituent particles. In the case under consideration, these are conformally coupled scalar fields. 
In other words, exactly the particles flowing through the banana loops. 
Higher-order corrections will modify the free theory scaling dimension, shifting it with functions of couplings which have been integrated out. 
These corrections lead to an \textit{anomalous dimension}, and they are turning the bananas into genuine unparticles. 
An explicit manifestation of this phenomenon would require the computation of all higher loops which falls outside the purview of this article. 

\section{Cosmological Differential Equations}
\label{sec:DE_eqns}

The simple integral representation \eqref{def:twi_int} can be exploited to derive a system of differential equations for generic $\epsilon_1$ and $\epsilon_2$ \cite{Pimentel:2025rds}.  
This so-called ``Pfaffian system" can be thought of as the auxiliary matrix of the two partial differential equations (PDEs): 
\begin{align}
    K_{\epsilon_1,\epsilon_2} &\bullet I_{++} = \left(D_{X_1}-D_{X_2}\right) I_{++} = 0, \label{eq:boosts} \\
    T_{\epsilon_1} &\bullet I_{++} = \left((X_1+X_2)\partial_{X_1}\partial_{X_2} - \epsilon_1 (\partial_{X_1} + \partial_{X_2})\right) I_{++} = 0, \label{eq:time_trans}
\end{align}
where $D_X \equiv (X^2-1)\partial_X^2 - 2(\epsilon_1+\epsilon_2)X\partial_X$. 
Here the twisted parameters are 
\begin{align}
    \epsilon_1 = (1+\delta)(2-n(1+\epsilon))-1, \quad \epsilon_2 = n(1+\epsilon)-2-\epsilon. 
    \label{def:twi_para_full}
\end{align}
The operator $K_{\epsilon_1,\epsilon_2}$ in \eqref{eq:boosts} can be regarded as an $\epsilon_1$ and $\epsilon_2$ deformation of the de Sitter boost \cite{Arkani-Hamed:2015bza,Arkani-Hamed:2018kmz} in a $D$-dimensional FRW spacetime. 
The $T_{\epsilon_1}$ operator in \eqref{eq:time_trans} has the interpretation of a time-translation operator in the same spacetime. 
These two operators form a so-called ``annihilating ideal" (or ``annihilator") in the rational Weyl algebra. The (holonomic) functions that are annihilated by the operators in this ideal form a four-dimensional solution space. 
We would like to refer to \cite{Fevola:2024nzj} for a comprehensive introduction to D-ideals in cosmology. 
We denote the ideal with fixed $\epsilon_1$ and $\epsilon_2$ by $\mathcal{I}_{\epsilon_1,\epsilon_2}$. 
Notice that the $(+-)$-sector, $I_{+-}$, also satisfies the boost invariance equation \eqref{eq:boosts}. 
The analogue of $T_{\epsilon_1}$ for $I_{+-}$ is obtained by simply flipping the sign of $X_1$ and $\partial_{X_1}$ in $T_{\epsilon_1}$.

Since we are mostly interested in the inflationary scenario, we would like to choose $\delta=0$, corresponding to a $D$-dimensional de Sitter spacetime.
This choice will simplify the twisted parameters to 
\begin{align}
    \epsilon_1 = 1-n(1+\epsilon), \quad \epsilon_2 = n(1+\epsilon)-2-\epsilon. 
    \label{def:twi_para_dS}
\end{align}
In general, IR-divergences occur when $\epsilon_1 \in \mathbb Z$ and UV-divergences occur if $\epsilon_2 \in \mathbb Z$. 
By fixing $\delta=0$, the long-range and short-range divergences are both regularized by the same parameter $\epsilon$. 
Within the family of banana-integrals parametrized by an integer number $n$, we need to distinguish between two cases: the tree-level exchange where $n=1$, and the $(n-1)$-loop diagrams for $n\geq2$. 
For $n=1$, the correlator is UV-finite and the $\epsilon$-poles can be associated with long-distance infinities. 
By summing the different sectors from the ``in-in" formalism, these IR-poles will cancel and this results in a finite answer, as explained in \cite{Arkani-Hamed:2018kmz}. 
On the other hand, since the two-point function of the composite operator $:\varphi^n:$ scales as $\sim1/|x|^{2n}$, the banana integrals with $n \geq 2$ converge faster at large distance than the tree-level integral. 
The $\epsilon$ poles in these loops are purely related to UV-divergences. 

\subsection{The Bubble Seed Function}
\label{sec:seed}
\vspace{\manspacing}

We describe a systematic method to generate all perturbative solutions in $\epsilon$ to the differential equations \eqref{eq:boosts} and \eqref{eq:time_trans} using only bootstrap information. 
The first step of this procedure is to solve for the bubble ($n=2$). 
Then, we will show in the next section how we can use shift operators to obtain higher loop ($n>2$) solutions from this bubble seed. 

In order to keep the small parameters in $\epsilon_1$ and $\epsilon_2$ generic, we will isolate $n$ and introduce two general multipliers for $\epsilon_1$ and $\epsilon_2$: 
\begin{equation}
    \epsilon_1=1-n+\lambda_1 \epsilon, \qquad \epsilon_2 = n - 2 + \lambda_2 \epsilon. 
    \label{def:twi_para_dS_multiplied}
\end{equation}
After applying shift operators to raise $n$, we can choose the values of $\lambda_1$ and $\lambda_2$ in order to match the dictionary \eqref{def:twi_para_dS}. Alternatively, more in the bootstrap spirit, we can keep them generic until renormalization where they will be determined by symmetry requirements.

The procedure starts by plugging in a general Laurent expansion 
\begin{align}
    \begin{aligned}
        I_{++}(X_1,X_2)&= \sum_{n=-m}^{\infty} \epsilon^{n} f^{(n)}(X_1,X_2) 
    \end{aligned}
\end{align}
into the differential equations. 
From \eqref{eq:boosts} and \eqref{eq:time_trans}, we can derive an ODE, associated to a differential operator $\text{O}_1$,
\begin{align}
    \begin{aligned}
        \text{O}_1 &= \partial_{X_2} \Bigg(\partial_{X_2}\Big((X_1+X_2)D_{X_2}\Big)- (\epsilon_1-1)(2\epsilon_1+2\epsilon_2+1)\Big) \\
        &+\epsilon_1\Big(-2D_{X_2} + \big((\epsilon_1+1)(X_1+X_2)+2\epsilon_2X_1-2X_2\big)\partial_{X_2}\Bigg),
    \end{aligned}
\end{align}
that is more amenable to solve. 
If we plug in the tree-level ($n=1$) parametrization $\epsilon_1=-\epsilon,\ \epsilon_2=-1$, we will recover equation (3.45) in \cite{Arkani-Hamed:2023kig}.
We can solve this ODE perturbatively in $\epsilon$. 
At every order, we obtain an Ansatz, of which the four remaining integration constants will be functions of $X_1$ that can be further fixed by imposing:
\begin{enumerate}
    \item One of the PDEs from \eqref{eq:boosts} and \eqref{eq:time_trans}; 
    \item Absence of folded singularities (when we take $X_1\rightarrow 1$ or $X_2\rightarrow 1$); 
    \item $X_1\leftrightarrow X_2$ symmetry. 
\end{enumerate}
Applying this procedure at the lowest order, $\epsilon^{-m}$, we find that the most general expression satisfying our bootstrap constraints will be 
\begin{equation}
    f^{(-m)}(X_1,X_2)=\frac{C_0}{X_1+X_2}+C_1, 
    \label{eq:-m_order_sol}
\end{equation}
where $C_0$ and $C_1$ are constants of integration that cannot be fixed by any of the three requirements mentioned above. 
When we solve for order $\epsilon^{-m+1}$, the ``total energy" solution \eqref{eq:-m_order_sol} will act as a source, thus generating an inhomogeneous differential equation for $f^{(-m+1)}$. 
Without introducing any $\log(X_1-1)$ dependence, the final result can be brought to the form
\begin{align}
    \begin{aligned}
        &f^{(-m+1)}(X_1,X_2) = \frac{C_2}{X_1+X_2}+C_3 \\
        &+ C_0 \frac{1}{X_1+X_2} \left(2\lambda_1+\lambda_2 + 2\lambda_1\log(X_1+X_2) + \lambda_2\log(X_1+1) + \lambda_2\log(X_2+1) \right). 
        \label{eq:-m+1_order_sol}
    \end{aligned}
\end{align}
$C_0$ is the constant inherited from \eqref{eq:-m_order_sol}, while $C_2$ and $C_3$ are new constants of integration that are again not constrained by any of the three requirements.

Notice that at this point, the value of $m$, the lowest order in the Laurent expansion, is also arbitrary. 
From the perspective of the linear differential equations and the other conditions we impose, the overall scaling (including the power of $\epsilon$) is arbitrary. 
Information about the proportionality constant is generally provided by a boundary condition where we evaluate the solution at a certain point in kinematic space. 
However, we want to determine $m$ from a bootstrap perspective. 
Therefore, we introduce a new bootstrap rule: 
\begin{align*}
    \textit{all $\epsilon$ poles must be absorbable by local interactions. }
\end{align*}
From our known solutions \eqref{eq:-m_order_sol} and \eqref{eq:-m+1_order_sol}, we know that $f^{(-m+1)}$ contains logarithms of $(X_1+1)$ and $(X_2+1)$ whereas the kinematic dependence of $f^{(-m)}$ is just the ``total energy" $(X_1+X_2)$. 
Obviously, only when $m=1$ the required counterterm to renormalize the theory attains a local dependence on the kinematic variables given by $1/(X_1+X_2)$. 
As a byproduct of our new bootstrap requirement, $C_1$ needs to be zero as well. 
The specific choice of $C_0$ is merely a normalization, which we choose to be $1$ for simplicity. 

The remaining constants, $C_2$ and $C_3$, can be fixed by examining two other limits of the correlator \cite{Arkani-Hamed:2018kmz}. 
The constant $C_2$ can be determined by taking the flat space limit $X_1\rightarrow-X_2$ of the correlator and matching the limiting expression with the amplitude. 
Working with the \textit{modified minimal subtraction} scheme makes its precise value irrelevant. Let us use this freedom to conveniently set it to $-2\lambda_1-\lambda_2$. 

Finally, we expect the correlator to be free of the $Y$ singularity in the collapsed limit. 
The exchange momentum, $Y$, does not show up explicitly in our formulas since we are working with rescaled variables, e.g. $X_1 = \tilde X_1/Y$ where $\tilde X_1$ and $Y$ have a mass-dimension which equals to one. 
From dilation invariance, we know that the overall dimension of the correlator is minus one \cite{Arkani-Hamed:2018kmz}. 
This implies that the dimensionful correlator can be written as $Y^{-1} I_{++}$. 
Multiplying every term in $I_{++}$ by $Y^{-1}$ and reinstating the dimensionful ``energies", the dimensionful correlator acquires the following schematic form:
\begin{equation*}
    \frac{A}{\tilde X_1+\tilde X_2} + \frac{C_3}{Y}.
\end{equation*}
The last term contains an unphysical $Y$ pole forcing $C_3$ to vanish.

In conclusion, to be fully consistent with all the bootstrap requirements, the constants are fixed to be
\begin{align}
    &C_0 = 1, & &C_1=0, & &C_2=-2\lambda_1-\lambda_2, & &C_3=0. 
\end{align}
Here $\lambda_1$ and $\lambda_2$ can be taken from \eqref{def:twi_para_dS} which would imply that $\lambda_1=-2$ and $\lambda_1=1$. 
However, we will argue in the next section how the ratio of these two values is dictated by boost invariance of the renormalized correlator in de Sitter. 
In conclusion, we arrive at the following expression for connected part of the seed-correlator: 
\begin{align}
    \begin{aligned}
        I_{++}^{(n=2)}(X_1,X_2) &= \frac{1}{\epsilon}\frac{1}{X_1+X_2} \\
        &+\frac{1}{X_1+X_2}\left(\lambda_2 \log (X_1+1)+\lambda_2\log (X_2+1)-2\lambda_1\log (X_1+X_2)\right). 
        \label{eq:seed}
    \end{aligned}
\end{align}
Before discussing renormalization, we will use the expression above to generate the perturbative expansions in $\epsilon$ for all $n$-loops. 

\subsection{Shift Operator and Explicit Solutions}
\label{sec:shift}
\vspace{\manspacing}

From the initial definition of the twisted parameters in de Sitter space \eqref{def:twi_para_dS}, we can see that shifting $n\mapsto n+1$ corresponds to lowering $\epsilon_1 \mapsto\epsilon_1-(1+\epsilon)$ and raising $\epsilon_2\mapsto\epsilon_2+(1+\epsilon)$. 
We have defined \eqref{def:twi_para_dS_multiplied} to keep the $\epsilon$ terms in $\epsilon_1$ and $\epsilon_2$ generic in the seed solution. 
These two parameters $\lambda_1$ and $\lambda_2$ allow us to specify $\epsilon_1$ and $\epsilon_2$ \emph{after} we act sufficiently many times with the shift operator.  

By employing partial fractions and integration by parts (IBP), the shift operator can be derived from the integral representation \eqref{def:twi_int}: 
\begin{equation}
    \sigma_+ \equiv -1 + \frac{1}{\epsilon_1}\frac{X_1^2-1}{X_1+X_2}\partial_{X_1}+\frac{1}{\epsilon_1}\frac{X_2^2-1}{X_1+X_2}\partial_{X_2}. 
    \label{def:shift_operator}
\end{equation}
The $\epsilon_1$ value is as defined in \eqref{def:twi_para_dS_multiplied}. 
A detailed derivation can be found in Appendix \ref{app:shift_operator}. 
The application of this shift operator to the correlator can be written as
\begin{equation}
    \sigma_+ \bullet I_{++}^{(\epsilon_1,\epsilon_2)} = I_{++}^{(\epsilon_1-1,\epsilon_2+1)}, 
    \label{def:shift_relation}
\end{equation}
while satisfying the right operator action: 
\begin{align}
    \sigma_+ \bullet K_{\epsilon_1,\epsilon_2} - K_{\epsilon_1-1,\epsilon_2+1} \bullet \sigma_+ &= \mathcal{K}_{\epsilon_1,\epsilon_2} \in \mathcal{I}_{\epsilon_1,\epsilon_2}, \label{eqn:commutation_K} \\
    \sigma_+ \bullet T_{\epsilon_1} - T_{\epsilon_1-1} \bullet \sigma_+ &= \mathcal{T}_{\epsilon_1} \in \mathcal{I}_{\epsilon_1,\epsilon_2}. \label{eqn:commutation_T}
\end{align}
These two equations \eqref{eqn:commutation_K} and \eqref{eqn:commutation_T} can be used to show the validity of \eqref{def:shift_relation}. 
Suppose that you are given a function $\psi_{\epsilon_1,\epsilon_2}$ living in the solution space of $\mathcal{I}_{\epsilon_1,\epsilon_2}$. 
Define $\psi_{\epsilon_1-1,\epsilon_2+1}$ as $\sigma_+ \bullet \psi_{\epsilon_1,\epsilon_2}$. 
Let's now act with both sides of \eqref{eqn:commutation_K} on $\psi_{\epsilon_1,\epsilon_2}$. 
Since both $K_{\epsilon_1,\epsilon_2}$ and $\mathcal{K}_{\epsilon_1,\epsilon_2}$ are elements of the ${\epsilon_1,\epsilon_2}$ annihilator, they kill $\psi_{\epsilon_1,\epsilon_2}$. 
This leaves only 
\begin{align}
    K_{\epsilon_1-1,\epsilon_2+1}\bullet \sigma_+ \bullet \psi_{\epsilon_1,\epsilon_2} = 0, 
\end{align}
which by definition reduces to 
\begin{align}
    K_{\epsilon_1-1,\epsilon_2+1}\bullet \psi_{\epsilon_1-1,\epsilon_2+1} = 0. 
\end{align}
By repeating the same story for the time-translations $T_{\epsilon_1}$, we conclude that $\psi_{\epsilon_1-1,\epsilon_2+1}$ must be an element of the $\mathcal{I}_{\epsilon_1-1,\epsilon_2+1}$-solution space. 

To see an example in action, let's apply the shift operator to the seed \eqref{eq:seed}. 
The resulting $n=3$ solution is 
\begin{align}
    \begin{aligned}
        I_{++}^{(n=3)} &=\sigma_+ \bullet I_{++}^{(n=2)} =  -\frac{1+X_1 X_2}{\left(X_1+X_2\right){}^3 }\frac{1}{\epsilon} \\ 
        &+\frac{X_1^2 + X_2^2-4 X_1 X_2+4 X_1+4X_2 - 6 -4 \left(X_1 X_2+1\right)\log \left(\frac{(X_1+1)(X_2+1)}{(X_1+X_2)^3}\right)}{2 \left(X_1+X_2\right){}^3}.
    \end{aligned}
\end{align}
As we can already infer from the example above, there is an increasing complexity as we go up in $n$.

We want to quickly point out that there is a similar raising operator for the $(+-)$-sector $I_{+-}$, which is obtained by multiplying $\sigma_+$ with $-1$ and sending $X_2 \mapsto -X_2$. 

\subsection{Renormalization}
\label{sec:renorm}
\vspace{\manspacing}

In the example we are discussing, the number of external legs for the $n$-bananas is fixed to be four. 
Therefore, the relevant counterterms come from four-point contact interactions $\varphi^4$ and its higher-derivatives. 
For the $n=2$ case, we consider the simplest type of contact interaction $g\varphi^4$ in the Lagrangian. 
The four-point correlator for this contact diagram in a general $D$-dimensional FRW background is a simple time-integral,
\begin{equation}
    I_{c,++} \equiv \int_{-\infty}^0 \frac{\dint \eta}{\eta^{\epsilon_1+1}} e^{i\eta X} \propto X^{\epsilon_1}, 
\end{equation}
where $\epsilon_1$ is again $-1+\lambda_1\epsilon$ as defined in \eqref{def:twi_para_dS_multiplied} and $X$ plays the role of a single vertex ``total energy".  
Alternatively, the contact diagram can be bootstrapped as done in \cite{Arkani-Hamed:2018kmz}.
Expanding in terms of $\epsilon$, we can see that the contact diagram is now corrected by orders of $\epsilon$: 
\begin{align}
    I_c = \frac{1}{X} + \frac{\lambda_1\epsilon}{X}\log(X) + O(\epsilon^2).
\end{align}
To cancel the UV-divergence of the loop, the contact correlator that we add, needs to be proportional to $1/\epsilon$. 
From the perspective of the Lagrangian, this corresponds to a counterterm of the form $\delta g=-1/\epsilon$. By adding the counterterm to the bare result, we find the renormalized bubble  
\begin{equation}
    \bar{I}^{(n=2)}_{++} \equiv I^{(n=2)}_{++} - \frac{1}{\epsilon} \, I_c = \frac{1}{X_1+X_2} \left(  
    \lambda_2\log(X_1+1) + \lambda_2\log(X_2+1) + \lambda_1 \log(X_1+X_2)\right).
\end{equation}
By construction, the result is finite when $\epsilon \rightarrow 0$ which allows us to act with a true de Sitter boost operator $K_{-1,0}$ on the solution. 
It turns out that this selects a unique combination of the parameters $\lambda_1$ and $\lambda_2$, which is $\lambda_1=-2\lambda_2$. 
The final renormalized result for the bubble is 
\begin{equation}
    \bar{I}^{(n=2)}_{++} = \frac{1}{X_1+X_2} \left(  
    \log(X_1+1) + \log(X_2+1) -2 \log(X_1+X_2)\right).
\end{equation}
This is coincidentally also the natural ratio of the $\lambda$'s from \eqref{def:twi_para_dS}. 
Our choice to renormalize the theory with local counterterms restricts the small parameters appearing in $\epsilon_1$ and $\epsilon_2$, and therefore the specific regularization scheme that we ought to use. 
This is exactly the reason why Weinberg's ``unphysical" logarithm correction to the power spectrum in inflation \cite{Weinberg:2005vy,Senatore:2009cf} cannot be renormalized by a local counterterm. 

In contrast to the preservation of boost invariance for the renormalized $\bar{I}_{++}^{(n=2)}$, the time-translation differential equation \eqref{eq:time_trans}, is not solved by this renormalized bubble. 
Instead, it picks up a source and becomes an inhomogeneous differential equation: 
\begin{equation}
    T_{-1} \bullet \bar{I}^{(n=2)}_{++} = \frac{2}{(X_1+X_2)^2}. 
\end{equation}
Notice that while the bare loop $I_{++}^{(n=2)}$, is annihilated by $T_{-1-2\epsilon}$, the counterterm is not. 
In fact, the source term is produced exactly by $T_{-1-2\epsilon}$ acting on the counterterm. 
This suggests that all the other renormalized $n$-banana loops satisfy similar inhomogeneous differential equations, where the sources are determined by $T_{\epsilon_1}$ acting on the right counterterms. 

Although computing the complete tower of counterterms for $n$-banana loops is beyond the scope of this paper, the shift relation allows us to study the divergent part in more generality. 
First of all, since the shift relation does not contain any poles in $\epsilon$, we know that as long as we start with a simple pole $1/\epsilon$ at $n=2$, all higher order bananas will only contain a simple pole in $\epsilon$. 
Furthermore, since the dynamical structure of the divergent term is not affected by a choice of $\lambda_1$ and $\lambda_2$, they can be inferred from applying the shift operator $(n-2)$ times to the total energy pole: 
\begin{equation}
    I_{++}^{(n)} \ \propto \ (\sigma_+)^{n-2} \, \bullet \, I_{++}^{(n=2)} \ + \ O\left(\epsilon^0\right).
\end{equation}
From \cite{Arkani-Hamed:2018kmz}, we know that the general contact interactions and their higher derivatives can be written as a linear combination of the terms generated by applying $D_{X_1}$ repeatedly to the ``total energy" pole in de Sitter: 
\begin{equation}
    I_c(X_1,X_2) = \sum_{m=0}^\infty c_m \, D_{X_1}^{\, m} \bullet \frac{1}{X_1+X_2} \, .
\end{equation}
We have checked up to $n=7$ that all simple poles in $\epsilon$ for the banana loops come with a kinematic dependence that can be written as an $I_c$ with different coefficients $c_m$. 
To the extent that we are able to verify, the divergences of the banana loops are all absorbable into local counterterms. 

In a similar vein, we can learn about the analytic structure of any banana loop. 
Since $\sigma_+$ is a first order differential operator, the logarithms of the seed function can only be transformed into a rational function or a term proportional to itself by $n$ applications of $\sigma_+$. 
This means that the logarithms are completely fixed for all $n$ to be a certain combination of $\log(X_1+1)$, $\log(X_2+1)$ and $\log(X_1+X_2)$. 
In other words, the whole family of banana correlators shares the same branch point structure.  

\vspace{-.2cm}
\section{Conclusions and Future Work}
\label{sec:discussion}

\begin{figure}[h]
    \centering
    \includegraphics[width=0.6\linewidth]{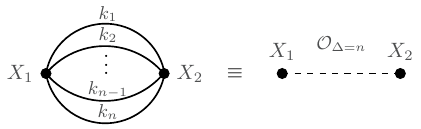}
    \caption{The new Feynman rule.}
    \label{fig:banana_rule}
\end{figure}

We demonstrated that banana loop diagrams with conformally coupled scalars are equivalent to tree-level unparticle exchanges with integer scaling dimension. Although we focused on the two-vertex diagram, our result is more general---we can introduce a ``new Feynman rule". For any banana loop composed of conformally coupled scalars connecting two vertices inside a multi-site correlator, we can replace the banana loop with an unparticle, as illustrated in Figure \ref{fig:banana_rule}. 
\vspace{-.2cm}
\vspace{-.2cm}

We have described a novel bootstrap method that only requires the differential equations and a few basic, physical constraints as input. 
The method can be used to derive a perturbative solution in the dim-reg parameter $\epsilon$ to any banana loop or unparticle exchange with any integer scaling dimension. 
The differential equations need to be solved for a only single seed, the bubble with $n=2$, after which one can use the shift relation to recursively obtain the tower of banana loops. 

The shift relations that we derived, enabled us to observe two general features of $n$-banana loops. 
First of all, their repeated action on the bubble has provided strong evidence that the UV-divergences can be absorbed into local counterterms for all $n$. 
Secondly, all correlators in this family share the same logarithmic branch points, at $X_1=-1,\ X_2=-1$ and $X_1=-X_2$. 

In the last section, we have detailed a renormalization procedure for this set of cosmological loops, emphasizing the subtleties that arise in an expanding spacetime. 
In particular, we have found that small deformations of the FRW scale factor and the dimension of the spacetime, controlled by $\lambda_1$ and $\lambda_2$ respectively, are intricately linked. 
Renormalizing the bubble with a local counterterm and subsequently requiring de Sitter boost invariance of the resulting finite expression, forces $\lambda_1$ and $\lambda_2$ to be of a specific ratio. 

There are many future directions to explore: 
\begin{itemize}
    \item A complete bootstrap of loops: the method to derive the perturbative results for the loop correlators in this paper heavily leaned towards a complete bootstrap. 
    The only essential ingredient that required knowledge about the integrals, is the set of differential equations. 
    There are essentially two avenues to search for a first-principles derivation of the differential equations: 
    \begin{enumerate}
        \item Deriving the $\epsilon_1$ and $\epsilon_2$ deformed operators directly. 
        A tantalizing possibility is to use the kinematic flow that was recently introduced in \cite{Arkani-Hamed:2023bsv}. 
        \item Computing the (inhomogeneous) differential equations in exact de Sitter in four dimensions ($\epsilon=0$) for the renormalized correlators. 
        An advantage here is that we already have one for free: the renormalized results were found to be invariant under the de Sitter boosts. 
    \end{enumerate}
    Once the differential equations have a symmetry underpinning them, it is worth looking into a derivation of the shift relations starting from the differential equations instead of the integral representation.
    \item We argued that the correspondence between the banana loops and unparticles arises when we turn off the self-interactions that bind the constituents of the unparticles together. 
    In this weakly coupled, perturbative scenario, the banana loops are the first order contributions to an unparticle exchange. 
    Although it would still be far from accessing the strongly coupled regime by means of a resummation, the story would be more complete by taking into account higher order corrections. 
    \item We only exchanged conformally coupled scalars in the banana loops. 
    Generalizing our results to include particles with arbitrary masses would be of paramount phenomenological importance. 
    In particular, it might be interesting to study the difference between heavy scalars from the \textit{principal series} and light scalars from the \textit{complementary series}. 
    Even for the simplest case, a bubble formed by scalars with generic mass in an FRW universe, the differential equations are not known. 
    It could be easier to start with a small massive deformation from the conformally coupled scalars and develop a perturbation theory with respect to $m/H$. 
    \item Along the same lines, it would be interesting to obtain more general results by means of differential operators. 
    Can we construct weight-shifting operators, where the spin and the mass of the external and exchanged particles are raised/lowered? 
    This would allow the results to be transformed into correlation functions of fields, such as the inflatons and gravitons, that are present during inflation. 
\end{itemize}
In conclusion, we have presented a novel approach to computing loop corrections to cosmological correlators, offering useful insights into their structure and leaving several promising directions for future research open.

\section*{Acknowledgements} 
Thanks to Craig Clark, Guilherme L. Pimentel, Jiaxin Qiao and Augusto Sagnotti for comments and insightful discussions. 
We are grateful to the reviewer for useful comments. 
We used the ``HolonomicFunctions" packages \cite{HolFun} to calculate Gröbner bases over the rational Weyl algebra. 
TW and CY are supported by Scuola Normale and by INFN (IS GSS-Pi). 
The research of TW and CY is moreover supported by the ERC (NOTIMEFORCOSMO, 101126304). 
Views and opinions expressed are, however, those of the author(s) only and do not necessarily reflect those of the European Union or the European Research Council Executive Agency. 
Neither the European Union nor the granting authority can be held responsible for them. 

\appendix
\section{Derivation of Shift Operator}
\label{app:shift_operator}

We start with the integral representation,
\begin{align}
    \begin{aligned}
        &\left\langle \frac{1}{x_1+x_2+X_1+X_2} \left(\frac{1}{x_1+t+X_1}+\frac{1}{x_2+t+X_2}\right) \right \rangle \\
        \equiv&\int_\Gamma \dint x_1 \, \dint x_2 \, \dint t\ \frac{(x_1 x_2)^{\epsilon_1} (t^2-1)^{\epsilon_2}}{x_1+x_2+X_1+X_2} \left(\frac{1}{x_1+t+X_1}+\frac{1}{x_2+t+X_2}\right), 
    \end{aligned}
\end{align}
with the twisted parameters specified to their $\delta=0$ de Sitter values given in \eqref{def:twi_para_dS}. 
The angled brackets denote an integral over the twisted cycle $\Gamma$. 
We define the shift operator through its action on the twisted parameters:
\begin{equation}
    \sigma_+:\quad \epsilon_1 \mapsto \epsilon_1 - 1, \ \epsilon_2 \mapsto \epsilon_2+1.
\end{equation} 
Simply operating with $\sigma_+$ on the integral, we obtain another twisted integral of which only the rational part is different, 
\begin{align}
    \begin{aligned}
        \sigma_+ \bullet \left\langle\frac{1}{x_1+x_2+X_1+X_2} \left(\frac{1}{x_1+t+X_1}+\frac{1}{x_2+t+X_2}\right)\right\rangle& \\
        = \left\langle\frac{t^2-1}{x_1 x_2}\frac{1}{x_1+x_2+X_1+X_2} \left(\frac{1}{x_1+t+X_1}+\frac{1}{x_2+t+X_2}\right)\right\rangle&. 
    \end{aligned}
    \label{eq:shifted}
\end{align}
This rational function can be decomposed into a sum of ``smaller'' rational functions. 
With smaller we mean rational functions with no linear dependence on $x_1$, $x_2$ or $t$ in the numerator and at most three terms in the denominator.
The first and second term in \eqref{eq:shifted} are related by a permutation of the external and internal variables $(X_1,x_1) \leftrightarrow (X_2,x_2)$. In the following, we will manipulate the first term and then use this permutation to symmetrize the result.
By means of partial fractions we write
\begin{align}
    \begin{aligned}
        &\frac{t^2-1}{x_1 x_2 \left(x_1+x_2+X_1+X_2\right) \left(x_1+t+X_1\right)} \\
        =& \frac{1}{x_2 \left(x_1+t+X_1\right)}+ \frac{X_1-X_2}{x_2 \left(x_1+x_2+X_1+X_2\right) \left(x_1+t+X_1\right)} \\
        -&\frac{1}{\left(x_1+x_2+X_1+X_2\right) \left(x_1+t+X_1\right)} +\frac{X_1^2-1}{X_1+X_2} \left(\frac{1}{x_1 x_2 \left(x_1+t+X_1\right)}\right. \\
        -&\frac{1}{x_1 \left(x_1+x_2+X_1+X_2\right) \left(x_1+t+X_1\right)}
        -\left. \frac{1}{x_2 \left(x_1+x_2+X_1+X_2\right) \left(x_1+t+X_1\right)}\right) \\
        -& \frac{1}{x_2 \left(x_1+x_2+X_1+X_2\right)} + \frac{t-X_1}{x_1 x_2 \left(x_1+x_2+X_1+X_2\right)}. 
    \end{aligned}
    \label{eq:partfrac}
\end{align}
We can simplify the decomposition by considering integration-by-parts (IBP) identities. 
Notice that the numerator of the last term in \eqref{eq:partfrac} still includes a linear $t$-dependence which can be eliminated by the IBP identity 
\begin{equation}
    \int \dint t \ (t^2-1)^{\epsilon_2}\ t\ =\ \frac{1}{2(\epsilon_2+1)}\int \dint \big[(t^2-1)^{\epsilon_2+1}\big]\ =\ 0\ .
    \label{eq:IBP1}
\end{equation} 
The IBP identity that encodes the symmetry of interchanging $x_1$ and $x_2$,
\begin{align}
    \begin{aligned}
        &\left\langle \frac{1}{x_1(X_1+X_2+x_1+x_2)} \right\rangle - \left\langle \frac{1}{x_2(X_1+X_2+x_1+x_2)} \right\rangle \\
        =& \frac{1}{\epsilon_1}\int \dint \left[ (x_1 x_2)^{\epsilon_1} (t^2-1)^{\epsilon_2} \ \dint \log(X_1+X_2+x_1+x_2) \wedge \dint t \right] = 0 \ ,
    \end{aligned}
\end{align}
reveals that the last line of \eqref{eq:partfrac} is proportional to $X_1-X_2$. 
Furthermore, the first term on the first line and the second term on the second line in \eqref{eq:partfrac} are proportional to respectively
\begin{align}
    &\int \dint \left[ (x_1 x_2)^{\epsilon_1} (t^2-1)^{\epsilon_2} \ \dint\log(X_1+x_1+t)\wedge\dint t \ \right] = 0, \\
    &\int \dint \left[(x_1 x_2)^{\epsilon_1} (t^2-1)^{\epsilon_2} \dint \log(x_1) \wedge \dint \log(X_1+x_1+t) \right] = 0. 
\end{align}
After symmetrizing, the shifted integral \eqref{eq:shifted} becomes 
\begin{align}
\begin{aligned}
    &\left\langle\frac{-1}{\left(x_1+x_2+X_1+X_2\right) \left(x_1+t+X_1\right)} + \frac{X_2^2-1}{X_1+X_2} \frac{-1}{x_2 \left(x_1+x_2+X_1+X_2\right) \left(x_1+t+X_1\right)} \right. 
    \\
    &\left.+ \frac{X_1^2-1}{X_1+X_2} \frac{-1}{x_1 \left(x_1+x_2+X_1+X_2\right) \left(x_1+t+X_1\right)} \right\rangle. 
\end{aligned}
\end{align}
Two IBP identities relate the last two terms to respectively the ${X_2}$ and ${X_1}$ derivatives of the original integral.
In conclusion, the shift operator is the following differential operator as defined in Section \ref{sec:shift}: 
\begin{equation}
    \sigma_+ = -1 + \frac{1}{\epsilon_1} \frac{X_1^2-1}{X_1+X_2} \partial_{X_1}+\frac{1}{\epsilon_1} \frac{X_2^2-1}{X_1+X_2} \partial_{X_2}. 
\end{equation}

\newpage

\addcontentsline{toc}{section}{Reference}
\bibliographystyle{utphys}
{\linespread{1.075}
\bibliography{reference}

\providecommand{\href}[2]{#2}\begingroup\raggedright\begin{thebibliography}{10}

\bibitem{Pimentel:2025rds}
G.~L. Pimentel and C.~Yang, ``{Strongly Coupled Sectors in Inflation: Gapless Theories and Unparticles},'' \href{http://arxiv.org/abs/2503.17840}{{\ttfamily arXiv:2503.17840 [hep-th]}}.

\bibitem{Green:2013rd}
D.~Green, M.~Lewandowski, L.~Senatore, E.~Silverstein, and M.~Zaldarriaga, ``{Anomalous Dimensions and Non-Gaussianity},'' \href{http://dx.doi.org/10.1007/JHEP10(2013)171}{{\em JHEP} {\bfseries 10} (2013) 171}, \href{http://arxiv.org/abs/1301.2630}{{\ttfamily arXiv:1301.2630 [hep-th]}}.

\bibitem{Kikuchi:2007az}
T.~Kikuchi and N.~Okada, ``{Unparticle Dark Matter},'' \href{http://dx.doi.org/10.1016/j.physletb.2008.06.021}{{\em Phys. Lett. B} {\bfseries 665} (2008) 186--189}, \href{http://arxiv.org/abs/0711.1506}{{\ttfamily arXiv:0711.1506 [hep-ph]}}.

\bibitem{Fox:2011pm}
P.~J. Fox, R.~Harnik, J.~Kopp, and Y.~Tsai, ``{Missing Energy Signatures of Dark Matter at the LHC},'' \href{http://dx.doi.org/10.1103/PhysRevD.85.056011}{{\em Phys. Rev. D} {\bfseries 85} (2012) 056011}, \href{http://arxiv.org/abs/1109.4398}{{\ttfamily arXiv:1109.4398 [hep-ph]}}.

\bibitem{CMS:2014jvv}
{\bfseries CMS} Collaboration, V.~Khachatryan {\em et~al.}, ``{Search for dark matter, extra dimensions, and unparticles in monojet events in proton\textendash{}proton collisions at $\sqrt{s} = 8$ TeV},'' \href{http://dx.doi.org/10.1140/epjc/s10052-015-3451-4}{{\em Eur. Phys. J. C} {\bfseries 75} no.~5, (2015) 235}, \href{http://arxiv.org/abs/1408.3583}{{\ttfamily arXiv:1408.3583 [hep-ex]}}.

\bibitem{Sannino:2009za}
F.~Sannino, ``{Conformal Dynamics for TeV Physics and Cosmology},'' {\em Acta Phys. Polon. B} {\bfseries 40} (2009) 3533--3743, \href{http://arxiv.org/abs/0911.0931}{{\ttfamily arXiv:0911.0931 [hep-ph]}}.

\bibitem{Cheung:2007zza}
K.~Cheung, W.-Y. Keung, and T.-C. Yuan, ``{Collider signals of unparticle physics},'' \href{http://dx.doi.org/10.1103/PhysRevLett.99.051803}{{\em Phys. Rev. Lett.} {\bfseries 99} (2007) 051803}, \href{http://arxiv.org/abs/0704.2588}{{\ttfamily arXiv:0704.2588 [hep-ph]}}.

\bibitem{Cheung:2007ap}
K.~Cheung, W.-Y. Keung, and T.-C. Yuan, ``{Collider Phenomenology of Unparticle Physics},'' \href{http://dx.doi.org/10.1103/PhysRevD.76.055003}{{\em Phys. Rev. D} {\bfseries 76} (2007) 055003}, \href{http://arxiv.org/abs/0706.3155}{{\ttfamily arXiv:0706.3155 [hep-ph]}}.

\bibitem{Groote:2005ay}
S.~Groote, J.~G. Korner, and A.~A. Pivovarov, ``{On the evaluation of a certain class of Feynman diagrams in x-space: Sunrise-type topologies at any loop order},'' \href{http://dx.doi.org/10.1016/j.aop.2006.11.001}{{\em Annals Phys.} {\bfseries 322} (2007) 2374--2445}, \href{http://arxiv.org/abs/hep-ph/0506286}{{\ttfamily arXiv:hep-ph/0506286}}.

\bibitem{Bzowski:2013sza}
A.~Bzowski, P.~McFadden, and K.~Skenderis, ``{Implications of conformal invariance in momentum space},'' \href{http://dx.doi.org/10.1007/JHEP03(2014)111}{{\em JHEP} {\bfseries 03} (2014) 111}, \href{http://arxiv.org/abs/1304.7760}{{\ttfamily arXiv:1304.7760 [hep-th]}}.

\bibitem{Mishnyakov:2023wpd}
V.~Mishnyakov, A.~Morozov, and P.~Suprun, ``{Position space equations for banana Feynman diagrams},'' \href{http://dx.doi.org/10.1016/j.nuclphysb.2023.116245}{{\em Nucl. Phys. B} {\bfseries 992} (2023) 116245}, \href{http://arxiv.org/abs/2303.08851}{{\ttfamily arXiv:2303.08851 [hep-th]}}.

\bibitem{Cacciatori:2023tzp}
S.~L. Cacciatori, H.~Epstein, and U.~Moschella, ``{Banana integrals in configuration space},'' \href{http://dx.doi.org/10.1016/j.nuclphysb.2023.116343}{{\em Nucl. Phys. B} {\bfseries 995} (2023) 116343}, \href{http://arxiv.org/abs/2304.00624}{{\ttfamily arXiv:2304.00624 [hep-th]}}.

\bibitem{Duhr:2025ppd}
C.~Duhr and S.~Maggio, ``{Feynman integrals, elliptic integrals and two-parameter K3 surfaces},'' \href{http://arxiv.org/abs/2502.15326}{{\ttfamily arXiv:2502.15326 [hep-th]}}.

\bibitem{Duhr:2025tdf}
C.~Duhr, ``{Modular forms for three-loop banana integrals},'' \href{http://arxiv.org/abs/2502.15325}{{\ttfamily arXiv:2502.15325 [hep-th]}}.

\bibitem{Chowdhury:2023arc}
C.~Chowdhury, A.~Lipstein, J.~Mei, I.~Sachs, and P.~Vanhove, ``{The Subtle Simplicity of Cosmological Correlators},'' \href{http://arxiv.org/abs/2312.13803}{{\ttfamily arXiv:2312.13803 [hep-th]}}.

\bibitem{Cacciatori:2024zrv}
S.~L. Cacciatori, H.~Epstein, and U.~Moschella, ``{Loops in de Sitter space},'' \href{http://dx.doi.org/10.1007/JHEP07(2024)182}{{\em JHEP} {\bfseries 07} (2024) 182}, \href{http://arxiv.org/abs/2403.13145}{{\ttfamily arXiv:2403.13145 [hep-th]}}.

\bibitem{Cacciatori:2024zbe}
S.~L. Cacciatori, H.~Epstein, and U.~Moschella, ``{Loops in anti de Sitter space},'' \href{http://dx.doi.org/10.1007/JHEP08(2024)109}{{\em JHEP} {\bfseries 08} (2024) 109}, \href{http://arxiv.org/abs/2403.13142}{{\ttfamily arXiv:2403.13142 [hep-th]}}.

\bibitem{Qin:2023nhv}
Z.~Qin and Z.-Z. Xianyu, ``{Nonanalyticity and on-shell factorization of inflation correlators at all loop orders},'' \href{http://dx.doi.org/10.1007/JHEP01(2024)168}{{\em JHEP} {\bfseries 01} (2024) 168}, \href{http://arxiv.org/abs/2308.14802}{{\ttfamily arXiv:2308.14802 [hep-th]}}.

\bibitem{Qin:2024gtr}
Z.~Qin, ``{Cosmological Correlators at the Loop Level},'' \href{http://arxiv.org/abs/2411.13636}{{\ttfamily arXiv:2411.13636 [hep-th]}}.

\bibitem{Benincasa:2024ptf}
P.~Benincasa, G.~Brunello, M.~K. Mandal, P.~Mastrolia, and F.~Vaz\~ao, ``{On one-loop corrections to the Bunch-Davies wavefunction of the universe},'' \href{http://arxiv.org/abs/2408.16386}{{\ttfamily arXiv:2408.16386 [hep-th]}}.

\bibitem{Maldacena:2002vr}
J.~M. Maldacena, ``{Non-Gaussian features of primordial fluctuations in single field inflationary models},'' \href{http://dx.doi.org/10.1088/1126-6708/2003/05/013}{{\em JHEP} {\bfseries 05} (2003) 013}, \href{http://arxiv.org/abs/astro-ph/0210603}{{\ttfamily arXiv:astro-ph/0210603}}.

\bibitem{Senatore:2009cf}
L.~Senatore and M.~Zaldarriaga, ``{On Loops in Inflation},'' \href{http://dx.doi.org/10.1007/JHEP12(2010)008}{{\em JHEP} {\bfseries 12} (2010) 008}, \href{http://arxiv.org/abs/0912.2734}{{\ttfamily arXiv:0912.2734 [hep-th]}}.

\bibitem{Senatore:2012nq}
L.~Senatore and M.~Zaldarriaga, ``{On Loops in Inflation II: IR Effects in Single Clock Inflation},'' \href{http://dx.doi.org/10.1007/JHEP01(2013)109}{{\em JHEP} {\bfseries 01} (2013) 109}, \href{http://arxiv.org/abs/1203.6354}{{\ttfamily arXiv:1203.6354 [hep-th]}}.

\bibitem{Pimentel:2012tw}
G.~L. Pimentel, L.~Senatore, and M.~Zaldarriaga, ``{On Loops in Inflation III: Time Independence of zeta in Single Clock Inflation},'' \href{http://dx.doi.org/10.1007/JHEP07(2012)166}{{\em JHEP} {\bfseries 07} (2012) 166}, \href{http://arxiv.org/abs/1203.6651}{{\ttfamily arXiv:1203.6651 [hep-th]}}.

\bibitem{Arkani-Hamed:2015bza}
N.~Arkani-Hamed and J.~Maldacena, ``{Cosmological Collider Physics},'' \href{http://arxiv.org/abs/1503.08043}{{\ttfamily arXiv:1503.08043 [hep-th]}}.

\bibitem{Anninos:2014lwa}
D.~Anninos, T.~Anous, D.~Z. Freedman, and G.~Konstantinidis, ``{Late-time Structure of the Bunch-Davies De Sitter Wavefunction},'' \href{http://dx.doi.org/10.1088/1475-7516/2015/11/048}{{\em JCAP} {\bfseries 11} (2015) 048}, \href{http://arxiv.org/abs/1406.5490}{{\ttfamily arXiv:1406.5490 [hep-th]}}.

\bibitem{Salcedo:2022aal}
S.~A. Salcedo, M.~H.~G. Lee, S.~Melville, and E.~Pajer, ``{The Analytic Wavefunction},'' \href{http://dx.doi.org/10.1007/JHEP06(2023)020}{{\em JHEP} {\bfseries 06} (2023) 020}, \href{http://arxiv.org/abs/2212.08009}{{\ttfamily arXiv:2212.08009 [hep-th]}}.

\bibitem{Cespedes:2023aal}
S.~C\'espedes, A.-C. Davis, and D.-G. Wang, ``{On the IR divergences in de Sitter space: loops, resummation and the semi-classical wavefunction},'' \href{http://dx.doi.org/10.1007/JHEP04(2024)004}{{\em JHEP} {\bfseries 04} (2024) 004}, \href{http://arxiv.org/abs/2311.17990}{{\ttfamily arXiv:2311.17990 [hep-th]}}.

\bibitem{Baumgart:2019clc}
M.~Baumgart and R.~Sundrum, ``{De Sitter Diagrammar and the Resummation of Time},'' \href{http://dx.doi.org/10.1007/JHEP07(2020)119}{{\em JHEP} {\bfseries 07} (2020) 119}, \href{http://arxiv.org/abs/1912.09502}{{\ttfamily arXiv:1912.09502 [hep-th]}}.

\bibitem{Gorbenko:2019rza}
V.~Gorbenko and L.~Senatore, ``{$\lambda \phi^4$ in dS},'' \href{http://arxiv.org/abs/1911.00022}{{\ttfamily arXiv:1911.00022 [hep-th]}}.

\bibitem{Banks:1981nn}
T.~Banks and A.~Zaks, ``{On the Phase Structure of Vector-Like Gauge Theories with Massless Fermions},'' \href{http://dx.doi.org/10.1016/0550-3213(82)90035-9}{{\em Nucl. Phys. B} {\bfseries 196} (1982) 189--204}.

\bibitem{Georgi:2007ek}
H.~Georgi, ``{Unparticle physics},'' \href{http://dx.doi.org/10.1103/PhysRevLett.98.221601}{{\em Phys. Rev. Lett.} {\bfseries 98} (2007) 221601}, \href{http://arxiv.org/abs/hep-ph/0703260}{{\ttfamily arXiv:hep-ph/0703260}}.

\bibitem{Georgi:2007si}
H.~Georgi, ``{Another odd thing about unparticle physics},'' \href{http://dx.doi.org/10.1016/j.physletb.2007.05.037}{{\em Phys. Lett. B} {\bfseries 650} (2007) 275--278}, \href{http://arxiv.org/abs/0704.2457}{{\ttfamily arXiv:0704.2457 [hep-ph]}}.

\bibitem{Georgi:2009xq}
H.~Georgi and Y.~Kats, ``{Unparticle self-interactions},'' \href{http://dx.doi.org/10.1007/JHEP02(2010)065}{{\em JHEP} {\bfseries 02} (2010) 065}, \href{http://arxiv.org/abs/0904.1962}{{\ttfamily arXiv:0904.1962 [hep-ph]}}.

\bibitem{Grinstein:2008qk}
B.~Grinstein, K.~A. Intriligator, and I.~Z. Rothstein, ``{Comments on Unparticles},'' \href{http://dx.doi.org/10.1016/j.physletb.2008.03.020}{{\em Phys. Lett. B} {\bfseries 662} (2008) 367--374}, \href{http://arxiv.org/abs/0801.1140}{{\ttfamily arXiv:0801.1140 [hep-ph]}}.

\bibitem{Stephanov:2007ry}
M.~A. Stephanov, ``{Deconstruction of Unparticles},'' \href{http://dx.doi.org/10.1103/PhysRevD.76.035008}{{\em Phys. Rev. D} {\bfseries 76} (2007) 035008}, \href{http://arxiv.org/abs/0705.3049}{{\ttfamily arXiv:0705.3049 [hep-ph]}}.

\bibitem{Hogervorst:2021uvp}
M.~Hogervorst, J.~Penedones, and K.~S. Vaziri, ``{Towards the non-perturbative cosmological bootstrap},'' \href{http://dx.doi.org/10.1007/JHEP02(2023)162}{{\em JHEP} {\bfseries 02} (2023) 162}, \href{http://arxiv.org/abs/2107.13871}{{\ttfamily arXiv:2107.13871 [hep-th]}}.

\bibitem{DiPietro:2021sjt}
L.~Di~Pietro, V.~Gorbenko, and S.~Komatsu, ``{Analyticity and unitarity for cosmological correlators},'' \href{http://dx.doi.org/10.1007/JHEP03(2022)023}{{\em JHEP} {\bfseries 03} (2022) 023}, \href{http://arxiv.org/abs/2108.01695}{{\ttfamily arXiv:2108.01695 [hep-th]}}.

\bibitem{Loparco:2023rug}
M.~Loparco, J.~Penedones, K.~Salehi~Vaziri, and Z.~Sun, ``{The K{\"a}ll{\'e}n-Lehmann representation in de Sitter spacetime},'' \href{http://dx.doi.org/10.1007/JHEP12(2023)159}{{\em JHEP} {\bfseries 12} (2023) 159}, \href{http://arxiv.org/abs/2306.00090}{{\ttfamily arXiv:2306.00090 [hep-th]}}.

\bibitem{Marolf:2010zp}
D.~Marolf and I.~A. Morrison, ``{The IR stability of de Sitter: Loop corrections to scalar propagators},'' \href{http://dx.doi.org/10.1103/PhysRevD.82.105032}{{\em Phys. Rev. D} {\bfseries 82} (2010) 105032}, \href{http://arxiv.org/abs/1006.0035}{{\ttfamily arXiv:1006.0035 [gr-qc]}}.

\bibitem{Fitzpatrick:2011hu}
A.~L. Fitzpatrick and J.~Kaplan, ``{Analyticity and the Holographic S-Matrix},'' \href{http://dx.doi.org/10.1007/JHEP10(2012)127}{{\em JHEP} {\bfseries 10} (2012) 127}, \href{http://arxiv.org/abs/1111.6972}{{\ttfamily arXiv:1111.6972 [hep-th]}}.

\bibitem{Rychkov:2016iqz}
S.~Rychkov, \href{http://dx.doi.org/10.1007/978-3-319-43626-5}{{\em {EPFL Lectures on Conformal Field Theory in D\ensuremath{>}= 3 Dimensions}}}.
\newblock SpringerBriefs in Physics. Springer Cham, 1, 2016.
\newblock \href{http://arxiv.org/abs/1601.05000}{{\ttfamily arXiv:1601.05000 [hep-th]}}.

\bibitem{Arkani-Hamed:2023kig}
N.~Arkani-Hamed, D.~Baumann, A.~Hillman, A.~Joyce, H.~Lee, and G.~L. Pimentel, ``{Differential Equations for Cosmological Correlators},'' \href{http://arxiv.org/abs/2312.05303}{{\ttfamily arXiv:2312.05303 [hep-th]}}.

\bibitem{Bautista:2019qxj}
T.~Bautista and H.~Godazgar, ``{Lorentzian CFT 3-point functions in momentum space},'' \href{http://dx.doi.org/10.1007/JHEP01(2020)142}{{\em JHEP} {\bfseries 01} (2020) 142}, \href{http://arxiv.org/abs/1908.04733}{{\ttfamily arXiv:1908.04733 [hep-th]}}.

\bibitem{Collins:1984xc}
J.~C. Collins, \href{http://dx.doi.org/10.1017/9781009401807}{{\em {Renormalization : An Introduction to Renormalization, the Renormalization Group and the Operator-Product Expansion}}}, vol.~26 of {\em Cambridge Monographs on Mathematical Physics}.
\newblock Cambridge University Press, Cambridge, 1984.

\bibitem{Iliesiu:2018fao}
L.~Iliesiu, M.~Kolo\u{g}lu, R.~Mahajan, E.~Perlmutter, and D.~Simmons-Duffin, ``{The Conformal Bootstrap at Finite Temperature},'' \href{http://dx.doi.org/10.1007/JHEP10(2018)070}{{\em JHEP} {\bfseries 10} (2018) 070}, \href{http://arxiv.org/abs/1802.10266}{{\ttfamily arXiv:1802.10266 [hep-th]}}.

\bibitem{Karlsson:2021duj}
R.~Karlsson, A.~Parnachev, and P.~Tadi\'c, ``{Thermalization in large-N CFTs},'' \href{http://dx.doi.org/10.1007/JHEP09(2021)205}{{\em JHEP} {\bfseries 09} (2021) 205}, \href{http://arxiv.org/abs/2102.04953}{{\ttfamily arXiv:2102.04953 [hep-th]}}.

\bibitem{Arkani-Hamed:2018kmz}
N.~Arkani-Hamed, D.~Baumann, H.~Lee, and G.~L. Pimentel, ``{The Cosmological Bootstrap: Inflationary Correlators from Symmetries and Singularities},'' \href{http://dx.doi.org/10.1007/JHEP04(2020)105}{{\em JHEP} {\bfseries 04} (2020) 105}, \href{http://arxiv.org/abs/1811.00024}{{\ttfamily arXiv:1811.00024 [hep-th]}}.

\bibitem{Fevola:2024nzj}
C.~Fevola, G.~L. Pimentel, A.-L. Sattelberger, and T.~Westerdijk, ``{Algebraic Approaches to Cosmological Integrals},'' \href{http://arxiv.org/abs/2410.14757}{{\ttfamily arXiv:2410.14757 [math.AG]}}.

\bibitem{Weinberg:2005vy}
S.~Weinberg, ``{Quantum contributions to cosmological correlations},'' \href{http://dx.doi.org/10.1103/PhysRevD.72.043514}{{\em Phys. Rev. D} {\bfseries 72} (2005) 043514}, \href{http://arxiv.org/abs/hep-th/0506236}{{\ttfamily arXiv:hep-th/0506236}}.

\bibitem{Arkani-Hamed:2023bsv}
N.~Arkani-Hamed, D.~Baumann, A.~Hillman, A.~Joyce, H.~Lee, and G.~L. Pimentel, ``{Kinematic Flow and the Emergence of Time},'' \href{http://arxiv.org/abs/2312.05300}{{\ttfamily arXiv:2312.05300 [hep-th]}}.

\bibitem{HolFun}
C.~Koutschan, {\em \href{http://www.koutschan.de/publ/Koutschan09/thesisKoutschan.pdf}{Advanced Applications of the Holonomic Systems Approach}}.
\newblock PhD thesis, RISC, Johannes Kepler University, Linz, September, 2009.

\end{thebibliography}\endgroup
}

\end{document}